# Comparison of two models of a Double Inlet Miniature Pulse Tube Refrigerator : Part B Electrical Analogy

**Yannick BAILLY, Philippe NIKA**

Centre de Recherche sur les Ecoulements, Surfaces et Transferts / UMR CNRS 6000
IGE Parc technologique, 2, avenue Jean Moulin, 90000 Belfort – France.
Tel. (33) 03 84 57 82 18 fax (33) 03 84 57 00 32 e-mail : bailly@ige.univ.fcomte.fr

**Abstract.**
The design of a Double Inlet Pulse Tube Refrigerator is investigated by means of an analogy with an electric circuit. The results obtained with the electric scheme are compared with both those of the thermodynamic model (Part A) and experiments. The basic formulation of equivalent electronic components is discussed and a few improvements are proposed to adjust the theoretical expressions of the electrical impedances of capillaries and regenerator in order to consider additional effects such pressure drops due to geometrical singularities at different internal flow regimes. A simplified formulation for the regenerator inefficiency is deduced from considerations on its harmonic functioning. The main purpose of this analysis considers especially the design of miniature cryocoolers dedicated to electronic applications. These models are applied to a commercial miniature refrigerator. A discussion of their relevance is achieved and a few suggestions on the refrigerator design are proposed to improve the cooling production.

**Keywords :** miniature pulse tube refrigerator, comparative modeling, electrical analogy, regenerator efficiency, optimization and design.

## Nomenclature

**letters**

$c_v$ , $c_p$ : heat capacities at constant volume and at constant pressure

*C* : Hydraulic capacity

*d* : diameter

*e* *tube thickness*

*E* : regenerator efficiency

$\dot{H}$ : enthalpy flux

*I i* *electric current*

*Jj* *complex imaginary*

$K_k$ : dimensionless conductance

$K_0 = \dot{m}/\Delta P$ : pressure drop factor

x : spatial variable

*Y=1/Z* electrical admittance

Z electrical impedance

**subscripts**

a : ambient

a : perturbation amplitude first order

c : cold side of tube cp : compressor

g : gas

h : hot zone

j : indices

h : hot zone

$l$ : length

$L$ : Hydraulic inductance

$m$, $\dot{m}$ : gas mass , gas mass flow rate

$P$, $\overline{P}$ : pressure,average gas pressure

Q : heat flow

r : ideal gas constant or radius

$R$ : Hydraulic resistance

$Re$ : Reynolds number

$Re_h$ Porous Reynolds number

S : section

T : temperature

t : time

U gas velocity

V : volume

W : mechanical work of gas

o : 0 order

re : regenerator

s : regenerator matrix

t : tube

**Greeks**

$\varepsilon$ : regenerator void fraction

$\gamma = {c_p}/{c_v}$ isentropical coefficient of gas

$\omega$ : angular frequency

$\rho$ : density

$\mu$ : dynamical viscosity

$\Lambda$ : pressure drop factor

$\langle\ \rangle$ time averaged value

## 1. Introduction.

The Double Inlet Pulse Tube Refrigerator (DIPTR)[1,2,3] or the Multiple Inlet Pulse Tube Refrigerator (MIPTR)[4] constitutes a great improvement in the concept of mini cryocoolers. This variety of Pulse Tube refrigerator presents a by-pass capillary between the hot end and the regenerator, and another capillary [5] between the hot end and the gas reservoir. Many spatial and electronic applications require the oscillator to be separated from the rest of the device to avoid mechanical vibrations and magnetic fields. Therefore, a third capillary is used to link the compressor and the first exchanger situated before the regenerator (see Fig.1 in part A). For miniature systems the influence of those organs has to be carefully studied : first they introduce parasite dead volumes and secondly they induce undesirable phase lags between pressure and gas velocity in the tube.

During last twenty years many theoretical models have been achieved because there was and there is always a great demand in applications for design. Most of published models are inapplicable for reliable practical design, either they introduce great mathematical difficulties, either they need a super calculator. Reasonable simplifications seems to be necessary, and the more logical is to limit the studies to a one dimensional description of the problem, introducing eventually some corrections. In this spirit, employing average axial values for the physical parameters seems reasonable, provided to take care of various phases. Such idea, associated with sinus decomposition of periodic variable and the linear approximation for small amplitude is on the foundations of most comprehensible modeling works[6,7,8]. In spirit of more important popularization this point of view leads naturally to introduce the use of simple equivalent electric scheme. So because the described devices operate in periodic regime, their action can be assimilated to that of an electric quadripole constituted of resistances, capacities and inductances[9,10,11]. This remark is exploited in this second part of the research.

In the previous part A of this paper, a model based on a thermodynamic analysis allowed to explore the internal variations of the functioning parameters : pressures, mass flows rates and cooling production. In this second part, the authors propose a more convenient approach to study and optimize the design of miniature DIPTR. As explained, in this model, each part of the cooler is represented by a basic electrical component, so most calculus can be achieved by using any classical commercial circuit solver. The relevance of this technique is tested by comparing results with the previous model and with experiments. Thanks to the harmonic regime assumption, the second model becomes more practical and takes into account much more parameters than the thermodynamic model : flow singularities, flow regimes, regenerator efficiency.

## 1. Theoretical modeling of the DIPTR with an electrical analogy

This modeling principle developed by the authors is specially applicable to miniature systems and consists in establishing an analogy with electric circuits only composed of resistances, capacities and inductances. This basic idea is presented in [9] for the Orifice Pulse Tube Refrigerator (OPTR), it seems that it has been first introduced in France by researchers of the Ecole Nationale Supérieure [10,11]. Despite the previous part indicates that pressures variations are not purely sinusoidal, this assumption is admitted. In this analogy between hydrodynamic and electrical domains, pressure and mass flow rate are assimilated to tension and current respectively.

### 2.1 Capillaries and regenerator

In their study Huang and Chuang [9] have developed a linear analysis for capillaries and regenerator in a OPTR. Making considerations on the general continuity, momentum and energy equations they deduce a practical dynamical modeling of the connecting tube and of the regenerator and introduce their transfer functions. In fact, any quadripole with known transfer function can be can represented by a simplest Pi or T quadripole. For a geometrical PI quadripole, composed of two parallel admittance $Y_1$,$Y_3$ separated by an impedance $Z_2$ expressed by eq. (1) and eq.(2). As in Microsystems capacities are small the first order of the develop of the functions are sufficient.

$$Z_2 = Sinh\left(\sqrt{Z_j Y_j}\right)\sqrt{\frac{Z_j}{Y_j}} \quad \rightarrow_{ZY\rightarrow 0} \quad Z_j \tag{1}$$

$$\frac{1}{Y_1} = \frac{1}{Y_3} = \frac{Sinh\left(\sqrt{Z_j Y_j}\right)}{Cosh\left(\sqrt{Z_j Y_j}\right)-1}\sqrt{\frac{Y_j}{Z_j}} \quad \rightarrow_{ZY\rightarrow 0} \quad \frac{2}{Y_j} \tag{2}$$

Where $Z_j$ and $Y_j$ are the characteristics impedance and admittance of the system under consideration.

$$Z_j = R_j + j\omega L_j \qquad Y_j = -j\omega C \tag{3}$$

For a given volume V, containing a presumable perfect compressible flowing gas, one can define the hydraulic resistance as the inverse of the hydraulic conductance from Part A eq. 10 :

$$R_j = \frac{|\Delta P|_j}{\dot{m}_j} = \frac{1}{K_j} = \frac{2\Lambda R_e \mu l}{\pi \bar{\rho} d^4} \tag{3}$$

This resistance represents regular pressure drops ; the value of $\Lambda$ depends on the flow regime : laminar or turbulent (see Part A eq. 11).

The same volume presents also an inductance and a capacity given by eq. (5) and eq.(6) :

$$L_j = \frac{|\Delta P|_j}{\partial \dot{m}_j / \partial t} = \frac{4\,l}{\pi\, d^2} \tag{5}$$

$$C_j = \frac{dm}{dP} = \frac{V}{r\,T\,\gamma} \tag{6}$$

In fact ,as explained later, it is not convenient to neglect singular pressure drops at the inlet and outlet of conduits. Moreover, additional effects may be introduced by internal geometry particularities. Therefore, a supplementary corrective term is introduced in eq.1 to improve the resistances evaluations accuracy :

$$R_j = \frac{128\mu l}{\pi \rho d^4}\left(\frac{aRe^{1-c} + bRe}{64}\right) + \frac{2\left(0.16 + K_1\right)\mu Re}{\pi \rho d^3} \tag{7}$$

In the case of a junction between two tubes, the coefficient $K_1$= 1.4. In addition, in 1922, Schiller proposed the factor 0.16 to take into account of entrance effects. *a, b, c* are coefficient depending on the flow regime.

The expression of the regenerator resistance is deduced from eq. 22 to eq. 26 of Part A. The inductance $L_{re}$ of the regenerator can be neglected in front of $R_{re}$ owing to the ratio of its length to its diameter.

$$R_{re} = \frac{|\Delta P|_{re}}{\dot{m}_{re}} = \frac{1}{K_{re}} = \frac{\mu l_{re} r \overline{T}_{re}}{2\varepsilon d_h^2 S_{re} \overline{P}_{re}} \left(a_{re} + b_{re} Re_h\right) \qquad L_{re} \approx 0 \tag{8}$$

Equations. (1) & (2) are valid only in the case of isothermal connecting tube, in fact as soon explained in micro regenerators axial thermal gradients $\partial T_{t0} / \partial x$ are important and corrections $\psi$ and $B_T$, are introduced in the expressions of equivalent Pi quadripole impedance (see appendix 2).

$$Z_2 = \frac{Z Sinh\left(\sqrt{Z_{re} Y_{re} \Psi}\right)}{\sqrt{Z_{re} Y_{re} \Psi}} \tag{9}$$

$$\frac{1}{Y_1} = \frac{1}{Y_3} = \frac{Z_{re} Sinh\left(\sqrt{Z_{re} Y_{re} \Psi}\right)}{\sqrt{Z_{re} Y_{re} \Psi}\left(Cosh\left(\sqrt{Z_{re} Y_{re} \Psi}\right) - 1\right) + \frac{B_T l_{re}}{2} Sinh\left(\sqrt{Z_{re} Y_{re} \Psi}\right)} \tag{10}$$

$$\Psi = \left(1 - j\omega \frac{\tau\, r}{c_g}\right) \qquad B_t = -\frac{1}{\overline{T}_{g0}} (1 - j\omega\tau) \frac{\partial T_{t0}}{\partial x} \tag{11}$$

. $\tau$ is the complex time constant given by eq. (A11) depending of the porous matrix and the envelop thermal capacities and also of the thermal exchange coefficients. In fact three convection coefficients are introduced : the external coefficient with the surroundings $h_e$, the coefficient between the gas and the internal tube wall $h'$ and the volumetric coefficient $H$ between the gas and the porous matrix. The two last coefficients are generally ignored in literature in the case of harmonic regimes. Nevertheless, some formulations can be obtained in [12,13,14].

$$\frac{H}{\sigma_s} = h' = 0.33 \frac{\lambda_g}{d_h} Re_h^{0.67} \tag{12}$$

$Re_h$ is the porous Reynolds number defined in Part A eq.23.

2.2 Compressor

The compressor is assimilated to a current source[6] of intensity $i_{cp}$ with a parallel capacity $C_{sc}$ depending on its swept volume, the temperature, the angular pulsation and the compression polytropic coefficient $\kappa$.

$$i_{cp} = \omega \overline{P} \kappa C_{sc} \quad \text{where} \quad C_{sc} = \frac{V_{sc}}{2\kappa r T_{sc}} \tag{13}$$

2.3 Global electric scheme

The final electric diagram of a DIPTR is reported on fig.1; previous results are avalaible. Each capillary connection (L1, L2) consists of one quadripole ( geometry PI) constituted of two capacities, one inductance and one resistance (the resistance and the inertance are series connected and the capacities are parallel). The heat exchangers (E1, E2 and E3), the tube ( T ) and the reservoir are likened to volumes of fluid hence they are represented only by capacities in the electric scheme. Lastly, the regenerator ( R1) consists of a resistance and a capacity equally distributed on both sides of this resistance. The resolution of the circuit, as soon as all impedances are known, provides information on various tensions (pressures) and currents ( mass flow rates) with their phases.

At the design beginning, the mass flow rates are unknown (so idem for *Re* and $\Lambda$ ) so an iterative procedure is required to obtain correct impedance values of each organ and especially of the three capillaries and the regenerator : initial Reynolds numbers *Re* are presumed, then the calculations are achieved and new *Re* are calculated again and so on. A dozen of iterations are usually required to stabilize the results (relaxation factor is necessary).

2.4 Cooling production

The maximal cooling production is deduced from the enthalpy flux through the gas tube with eq. (14) :

$$\left\langle \dot{Q}_{c\,\max} \right\rangle = \frac{\omega c_{pg}}{2\pi} \int_{t_0}^{t_0+\frac{\pi}{\omega}} \dot{m}_c(t)\left(T_c - T_{lre}(t)\right)dt \tag{14}$$

The inlet temperature calculus at the cold zone $T_{lre}$ is similar to that exposed in Part A eq.47 applying the organization chart reported on fig.5 Part A. In this way, shuttle effects in the gas tube are considered.

Additional model improvement concerning the regenerator inefficiency can be introduced to take into account the thermal losses $\left\langle \dot{H}_{re} \right\rangle$ (see Part A.eq. 53.). Thanks to the electric analogy, the pressure and velocity amplitudes calculus are easier ; this permits to approximate $\left\langle \dot{H}_{re} \right\rangle$ by the following expression detailed in appendix 3 :

$$\left\langle \dot{H}_{re} \right\rangle \approx c_p \rho_{g0} S_{re} \frac{1}{2} \Re\left[ T_{ga} \widetilde{U}_a \right] = S_{re} \frac{1}{2} \Re\left[ \tau \left( j\omega p_a \widetilde{U}_a - c_p \rho_{g0} \frac{\left| U_a \right|^2}{\varepsilon} \frac{\partial T_{t0}}{\partial x} \right) \right] \qquad (15)$$

Equation 15 shows that the regenerator inefficiency results from the contribution of two phenomena : on the one hand, the thermal gradient $\frac{\partial T_{t0}}{\partial x}$ at the tube wall and on the other hand, the imperfect internal thermal exchanges trough the complex time constant $\tau$ given by eq. (A11). Perfect thermal exchanges posses infinity exchange coefficients and are characterized by a quasi null time constant.

Finally the net cooling production of the DIPTR remains as in Part A :

$$\left\langle \dot{Q}_{cnet} \right\rangle = \left\langle \dot{Q}_c \right\rangle - \left\langle \dot{Q}_{shut} \right\rangle - \left\langle \dot{H}_{reg} \right\rangle - \left\langle \dot{Q}_{loss} \right\rangle \qquad (16)$$

## 2. Comparison of the two models with experiments

Experimental tests are achieved on a commercial (U shaped) miniature DIPTR manufactured by THALES CRYOGENIE. Figure 2 represents a photography of this apparatus whose main characteristics are reported in Table 1. In order to limit external losses by convection or radiation, the cold zone of the cooler is protected by a cryostat under vacuum. A little thin film resistance is coated at the extremity of the cold zone to simulate the heat generation of an electronic component. The cooler is filled with helium. The pressure is about 13 bars ; the maximal compressor operating frequency is 27 Hz.

In a first step, to keep the same conditions for the two models (thermodynamic and electric analogy), the regenerator efficiency and singular pressure drops are not considered. Figure 3 shows the pressure amplitude estimations provided by these models and compares them with experimental results. With the theoretical values of dimensionless conductances Kk evaluated with Part A eq. 39, the pressure amplitudes are underestimated. In fact a decrease of reduced conductances permits to improve the accordance between theoretical and experimental results. This is easy explained by the additional singular pressure drops existing in the experimental device actually : holes, curvatures, couplings. Concerning the dimensionless cooling power (Part A eq. 38), the same phenomenon is observed on fig.4 : for example Kkr = 3, Kk1 = 10 and Kkc = 97.7 seam more adequate values than initial estimations Kkr = 4.29, Kk1 = 17.55 and Kkc = 97.7. The thermodynamic model using the empirical conductances values seems quite exact when applied to the current case. However, major inconveniences are to be noticed : very heavy algorithm developments, long calculus time, step by step method, numerical resolutions of differential equations, numerical roots and integration… In the same conditions, the electrical analogy also underestimates the pressures amplitudes and overestimates the cooling power of about 25%. But its easier use allows to increase the iterations number and finally to obtain better results in a short time. The observed discrepancy may also result of the assumption that physical parameters obey to perfect sinusoidal evolutions. Besides, the influence of internal heat transfers in the pulsed tube is ignored, other important uncertainties are introduced by considering a constant regenerator inefficiency and the compressor polytropic factor is arbitrary fixed ($\kappa$=1.4) in eq.12.

## 4. Parametrical study for optimization

### 4.1 Reservoir and by-pass capillary conductance

Figure 5 shows the reservoir junction capillary conductance influence on the refrigerator performances calculated with the two models and three examples of by-pass conductance Kk1 (the resistances and the regenerator efficiency are still supposed constant). Maximums are more pronounced with the thermodynamic model than with the electrical analogy. Nevertheless the global shapes of curves remains comparable and the maximums are reached for neighbor values of Kkr. If Kkr is too high, the cooling production strongly decreases ; this is due to the bad phase adaptation between pressure and mass flow rates and also to the reduction of the pressure amplitude. This observation emphasizes at low temperature (figure 5 corresponds to Tc/Th=0.75). Figure 6 gives an example of velocities and pressure histories, phase lags between the pressure and the mass flow rate through the warm and cold extremities clearly appear. In the plotted configuration, the phase lag is relatively low, favoring the cooling effects. These graphs indicates that the design of a DIPTR strongly depends on numerous parameters. In fact, for a given basic geometrical configuration, after the operating temperatures and frequency are fixed, the first step consists in choosing optimal values of the capillaries and especially the reservoir link (i.e. Kkr). In this initial process, all components of the refrigerator are supposed ideal. For example, fig.5 shows optimal values of Kkr in a range from about 1.5 to 3 witch also depend of the conductance Kk1 of the by-pass. Here, the best performances appear when Kk1=0 (no bypass) but this result has to be considered carefully because calculations are achieved with constant regenerator efficiency value E=0.95. In fact, this efficiency depends on the fluid flow rate through the regenerator so it is modified by Kk1, the role of the by-pass consisting in improving E by relieving the regenerator without affecting the pressure amplitude in the pulse tube.

After Kkr and Kk1 are chosen, the second step consists in designing the heat exchangers and the regenerator, the pressures and the mass flow rates being known approximately. Finally, designing a DIPTR is achieved by iterating this procedure and introducing successive improvements. Thus, the complete eq. 4 can be used in the electrical analogy (not possible with the thermodynamical model). Then conductances are auto-determined : Kkr=1.2, Kk1=6, Kkc = 27.5, but the calculation is achieved with *a, b,* c values corresponding to the case of a static hydrodynamic regime instead of a periodic regime. This procedure is illustrated on fig.7 : a comparison of the results of the electrical analogy with and without the iterative procedure used to determine the resistances and the regenerator efficiency is reported. In both cases, the minimal cold temperatures reached are approximately the same and the curves slopes are different from experiments revealing defaults in the thermal losses evaluation. Moreover, the maximal cooling production is better estimated by the iterative procedure because more physical phenomena are considered.

### 4.2 Compressor swept volume effect

The ratio $V_{sc}/V_t$ between the compressor volume and the tube volume (or total device volume) fixes internal pressure amplitudes so it dramatically influences the cooling power and consequently the minimal temperature limit. It seems obvious that lowest temperatures are reached at the end of the gas expansion : fig.8 shows an example of pressure and temperature histories. So, in this transformation, the expansion ratio should be as great as possible. However, for high values of $V_{sc}/V_t$ a great care has to be taken about gas displacements to prevent shuttle effects between the two exchangers (cold and warm zones). Besides, when $V_{sc}/V_t$ increases, the cooling power is improved but the compressor work Wcp is also greater so the refrigeration efficiency COP (see eq. 17) diminishes. Figure 9 presents the evolution of this theoretical global refrigeration efficiency versus the swept volume of the compressor. These results indicate an optimum range to select the swept volume as explained in previous remarks. The compressor used in our experiments has a swept volume of only 1.66 cm3 so it could be implemented by a factor two approximately as indicated on fig.9.

$$COP = \frac{\langle \dot{Q}_{cnet} \rangle}{\frac{1}{2}\Re\left[p_a \widetilde{U}_a\right]} \qquad (17)$$

### 4.3 Regenerator geometry

The last improvements of the model permit to study the influence of the regenerator geometry (case of a fixed regenerator volume). For example, fig. 10 presents the cooling power versus the regenerator length ; in this case, the diameter is adjusted to conserve a constant volume. Again, an optimum is observed. In fact two contrary effects takes place : in a long regenerator the efficiency is better and the axial thermal gradient is lower but pressure drops emphasizes.

Considering eq. 15, eq.18 allows to estimate the regenerator efficiency. Concerning the refrigerator used in experiments, the minimum value of 0.015 m given by fig.11 is respected so the efficiency of the regenerator is better than 0.95 (value initially used in models).

$$E = 1 - \frac{\langle \dot{H} \rangle}{\langle \dot{W}_{acou} \rangle} \text{ with } \langle \dot{W}_{acou} \rangle = \frac{1}{2} S \Re\left[p_a \widetilde{U}_a\right] \qquad (18)$$

### 5. Conclusions

The aim of the study was to popularize the theoretical modeling of a DIPTR and to permit practical design. Making this an inventory of the whole parameters influencing a DIPTR functioning was also performed . In this work, the authors also tried to take into account a maximum of phenomena in a way as more realistic as possible. Particularly a great attention was bring for the estimation of the gas tube extremities temperatures : for miniature device cold and warm zone are very close and losses by shuttle effect of the gas must be considered. Another progress in this study is to take into account the effect on thermal conduction along the wall of the regenerator envelop, such improvement is generally neglected in models. The influence of connecting tubes is also considered and important at this scale. Two theoretical models dedicated to the design of DIPTR at small scale have been presented. Both models allow to simulate experimental results with an accuracy about 20 %. The first model is very theoretical. It is directly inspired from basic laws of thermodynamics, fluid mechanics and heat transfer and includes several results from thermoacoustics. As a result, the calculus are still complex and its programming is more difficult so it not very easy to use. Therefore, the authors propose to build an alternative model easier to use, the basic idea consisting in introducing the principle of the electrical analogy. The first model seems to be more accurate in some cases but the estimations provided by the second one remains sufficient to design a DIPTR prototype. In fact the practicability of the electrical analogy permits to improve iterative procedures counterbalancing defaults due to simplifications.

The particular problems induced by miniature coolers are due to a general lack of knowledge concerning many phenomena in periodic regime that can not be neglected at small scale : pressure drop factors, thermal losses factors, compressor polytropic factor, average porosity of regenerator porous matrix. Therefore, the current modeling of micro-exchangers (and especially micro-regenerator) is not realistic enough comparing to experimental data. The same remarks remain valid about capillaries in periodic compressible flow regime so impedances calculations are not entirely validated yet.

In further researches, the authors intend to improve these models to integrate additional physical laws. A global optimization procedure is envisaged to estimate the more efficient device geometry. Moreover all further developments are reliant on experimental results. Thus, a specific instrumentation has to be developed especially dedicated to such micro thermal systems.

## Acknowledgements

This work is supported by the company THALES CRYOGENIE (France).

## Appendix 1: Thermal modeling of the regenerator in harmonic regime

The three averaged equations of energy [12,13] for the gas, the porous matrix and the regenerator envelop in harmonic regime in the axial direction are :

$$(1-\varepsilon)\frac{\partial \overline{T_s}}{\partial t} = \frac{H}{\rho_s c_s}\left(\overline{T_g} - \overline{T_s}\right) + \frac{\lambda_s^*}{\rho_s c_s}(1-\varepsilon)\frac{\partial^2 \overline{T_s}}{\partial x^2} \qquad \text{(A1)}$$

$$\frac{\partial \overline{T_t}}{\partial t} = \frac{\sigma_e h_e}{\rho_t c_t}\left(T_a - \overline{T_t}\right) + \frac{\sigma_2 h'}{\rho_t c_t}\left(\overline{T_g} - \overline{T_t}\right) + \frac{\lambda_t}{\rho_t c_t}\frac{\partial^2 \overline{T_t}}{\partial x^2} \qquad \text{(A2)}$$

$$\left(\varepsilon\frac{\partial \overline{T}_g}{\partial t}+\overline{U}\frac{\partial \overline{T}_g}{\partial x}\right)=\frac{H}{\rho_g c_g}\left(\overline{T}_s-\overline{T}_g\right)+\frac{\varepsilon\sigma_1 h'}{\rho_g c_g}\left(\overline{T}_t-\overline{T}_g\right)+\frac{\lambda_g^*}{\rho_g c_g}\varepsilon\frac{\partial^2 \overline{T}_g}{\partial x^2}+\frac{\varepsilon}{\rho_g c_g}\frac{\partial p}{\partial t}+\frac{\overline{U}}{\rho_g c_g}\frac{\partial p}{\partial x}$$

with $\sigma_s=(1-\varepsilon)\,{6}/{d_p}$ $\quad\sigma_1=\frac{2}{r_i}$ $\quad\sigma_2=\frac{2r_i}{e(2r_i+e)}$ $\quad\sigma_e=\frac{2(r_i+e)}{e(2r_i+e)}$ (A4)

$h_e$ is an external convection coefficient for the envelop, $h'$ an internal thermal exchange coefficient and $H$ a volumetric thermal coefficient between the gas and the matrix.

Average values and periodic amplitudes of each term are introduced according to :

$$p=p_0+p_a e^{j\omega t}+p_{2a}e^{2j\omega t};\ \rho_g=\rho_{g0}+\rho_{ga}e^{j\omega t}+\rho_{g2a}e^{2j\omega t}$$
$$\overline{T}_g=T_{g0}+T_{ga}e^{j\omega t}+T_{g2a}e^{2j\omega t};\ \overline{T}_s=T_{s0}+T_{sa}e^{j\omega t}+T_{s2a}e^{2j\omega t};\ \overline{T}_t=T_{t0}+T_{ta}e^{j\omega t}+T_{t2a}e^{2j\omega t} \quad \text{(A5)}$$

System for order 0, with hypothesis $\frac{\lambda_s^*}{\rho_s c_s}\frac{\partial^2 T_{s0}}{\partial x^2}$ neglected, gives:

$$p_0=cst,\quad \rho_{g0}=\frac{p_0}{rT_{g0}},\quad T_{g0}=T_{s0}=T_{t0} \quad \text{(A6)}$$

$$T_{t0}=T_a+\left(T_h-T_a\right)\frac{\exp\left(\sqrt{B}(l_{re}-x)\right)-\exp\left(-\sqrt{B}(l_{re}-x)\right)}{\exp\left(\sqrt{B}\,l_{re}\right)-\exp\left(-\sqrt{B}\,l_{re}\right)}+\left(T_c-T_a\right)\frac{\exp\left(\sqrt{B}\,x\right)-\exp\left(-\sqrt{B}\,x\right)}{\exp\left(\sqrt{B}\,l_{re}\right)-\exp\left(-\sqrt{B}\,l_{re}\right)}$$

where $B=\frac{h_e\sigma_e}{\lambda_t}$ (A7)

System for first order is :

$$j\omega c_g\rho_{g0}T_{ga}+c_g\rho_{g0}\frac{U_a}{\varepsilon}\frac{\partial T_{g0}}{\partial x}-\frac{H}{\varepsilon}\left(T_{sa}-T_{ga}\right)-\sigma_1 h'\left(T_{ta}-T_{ga}\right)-j\omega p_a=0$$
$$j\omega(1-\varepsilon)T_{sa}-\frac{H}{\rho_s c_s}\left(T_{ga}-T_{sa}\right)-\frac{\lambda_s^*}{\rho_s c_s}(1-\varepsilon)\frac{\partial^2 T_{sa}}{\partial x^2}=0 \quad \text{(A8)}$$
$$j\omega T_{ta}+\frac{\sigma_e h_e}{\rho_t c_t}T_{ta}-\frac{\sigma_2 h'}{\rho_t c_t}\left(T_{ga}-T_{tsa}\right)-\frac{\lambda_t}{\rho_t c_t}\frac{\partial^2 T_{ta}}{\partial x^2}=0$$

The resolution of this system for $\frac{\lambda_s^*}{\rho_s c_s}(1-\varepsilon)\frac{\partial^2 T_{sa}}{\partial x^2}$ and $\frac{\lambda_t}{\rho_t c_t\omega\, l_{re}^2}<<1$ ,gives the local temperature amplitude of the envelop, the gas and the matrix :

$$T_{ta}=\sigma_2 h'\frac{\left(j\omega p_a-c_g\rho_{g0}\frac{U_a}{\varepsilon}\frac{\partial T_{t0}}{\partial x}\right)}{C_{plx12}},\ T_{ga}=\frac{\tau}{c_g\rho_{g0}}\left(j\omega p_a-c_g\rho_{g0}\frac{U_a}{\varepsilon}\frac{\partial T_{t0}}{\partial x}\right)$$

$$T_{sa} = \frac{\tau}{c_g \rho_{g0}} \frac{\left( j\omega p_a - c_g \rho_{g0} \frac{U_a}{\varepsilon} \frac{\partial T_{t0}}{\partial x} \right)}{1 + j\omega \frac{\rho_s c_s}{H} (1-\varepsilon)} \tag{A9}$$

where

$$C_{plxden} = \sigma_1 h' + j\omega \left( c_g \rho_{g0} + \frac{1-\varepsilon}{\varepsilon} c_s \rho_s \frac{H}{H + j\omega c_s \rho_s (1-\varepsilon)} \right) \tag{A10}$$

$$C_{plx12} = \sigma_1 h' (\sigma_e h_e + j\omega \rho_t c_t) - \left(\rho_t c_t \omega^2 - j\omega(\sigma_e h_e + \sigma_2 h')\right) \left( c_g \rho_{g0} + \frac{1-\varepsilon}{\varepsilon} c_s \rho_s \frac{H}{H + j\omega c_s \rho_s (1-\varepsilon)} \right)$$

$$\tau = \rho_{g0} c_g \frac{1 + \frac{\sigma_1 \sigma_2 h'^2}{C_{plx12}}}{C_{plxden}} \tag{A11}$$

## Appendix 2: hydrodynamic modeling of the regenerator in harmonic regime

Under the hypothesis density and velocity are constant in a section, the one dimensional continuity and momentum equations are:

$$\frac{\partial \rho_g}{\partial t} + \frac{\rho_g}{\varepsilon} \frac{\partial \overline{U}}{\partial x} + \frac{\overline{U}}{\varepsilon} \frac{\partial \rho_g}{\partial x} = 0 \tag{A12}$$

$$\frac{\rho_g}{\varepsilon} \frac{\partial \overline{U}}{\partial t} = -\frac{\partial p}{\partial x} - \frac{\mu_g \overline{U}}{K_p} \tag{A13}$$

Permeability $K_p$ is expressed by mean of the pressure drop coefficient $\Lambda_{Re}$ and the Reynolds number

$$K_p = \frac{2\varepsilon D_h^2}{Re_h \Lambda_{Re}} \quad \text{with} \quad \Lambda_{Re} = \frac{a}{Re_h} + b \quad \text{and} \quad Re_h = \frac{\rho_g \overline{U} D_h}{\varepsilon \mu} \tag{A14}$$

Once again introducing average values and periodic amplitudes of each term :

$$\overline{U} = U_a e^{j\omega t} + U_{2a} e^{2j\omega t} \tag{A15}$$

One establish equations to be solved for the first order

$$j\omega \rho_{ga} + \frac{1}{\varepsilon} \left( \rho_{g0} \frac{\partial U_a}{\partial x} + U_a \frac{\partial \rho_{g0}}{\partial x} \right) = 0$$

$$j\omega \rho_{g0} \frac{U_a}{\varepsilon} + \frac{\partial p_a}{\partial x} + \frac{\mu}{K_p} U_a = 0 \qquad \text{A(16)}$$

$$p_a = r\rho_{g0} T_{ga} + r T_{g0} \rho_{ga}$$

Substituting solutions for temperatures (A11) amplitudes are :

$$U_a = -\frac{1}{Z_U}\frac{\partial p_a}{\partial x} \quad \rho_{ga} = -\frac{1}{Y_U Z_U p_0}\left(\rho_{g0}\frac{\partial^2 p_a}{\partial x^2} + \frac{\partial \rho_{g0}}{\partial x}\frac{\partial p_a}{\partial x}\right) \tag{A17}$$

With the complex impedance and admittance by unite length.:

$$Z_U = j\omega\frac{\rho_{g0}}{\varepsilon} + \frac{\mu}{K_p} \quad \text{and} \quad Y_U = -\frac{j\omega\varepsilon}{p_0} \tag{A18}$$

Following the resolution one obtain for the pressure, a wave equation (A19), associated with its boundaries conditions (A21). The coefficient $B_T$ and $C_t$ are x depending but average values are used.

$$\frac{\partial^2 p_a}{\partial x^2} + B_T\frac{\partial p_a}{\partial x} + C_T p_a = 0 \tag{A19}$$

$$B_T = \frac{1}{T_{g0}}(j\omega\tau - 1)\frac{\partial T_{t0}}{\partial x} \quad \text{and} \quad C_T = Z_U Y_U\left(1 - j\omega\frac{r}{c_g}\tau\right) \tag{A20}$$

$$x = 0 \qquad p_a(0) = p_{a0} \qquad \frac{\partial p_a}{\partial x}(0) = -Z_U U_{a0} \tag{A21}$$

So remembering that mass flow rate is $\dot{m}_a = \rho_{g0} S U_a$, one obtain in term of electrical analogy, the transfer function between pressure and mass flow rates :

$$\begin{bmatrix} p_{a0} \\ \dot{m}_{a0} \end{bmatrix} = Exp\left(\frac{B_T}{2}L\right)\begin{bmatrix} (-\frac{B_T}{\sqrt{\Delta}}Sinh\left(\frac{L\sqrt{\Delta}}{2}\right) + Cosh\left(\frac{L\sqrt{\Delta}}{2}\right)) & (\frac{2Z}{L\sqrt{\Delta}}Sinh\left(\frac{L\sqrt{\Delta}}{2}\right)) \\ (-\frac{2C_T L}{Z\sqrt{\Delta}}Sinh\left(\frac{L\sqrt{\Delta}}{2}\right) & (Cosh\left(\frac{L\sqrt{\Delta}}{2}\right) + \frac{B_T}{\sqrt{\Delta}}Sinh\left(\frac{L\sqrt{\Delta}}{2}\right)) \end{bmatrix}\begin{bmatrix} p_{aL} \\ \dot{m}_{aL} \end{bmatrix} \tag{A22}$$

$$C_T = \frac{ZY}{L^2}\left(1 - j\omega\tau\frac{r}{c_g}\right); \quad \Delta = B_T^2 - 4C_T \tag{A23}$$

Now, global impedance and admittance for mass flow rate are:

$$Z = \frac{L\mu}{K_p\bar{\rho}_{g0}S} + j\omega\frac{L}{\varepsilon S}; \quad Y = -\frac{LS\varepsilon}{r\bar{T}_{g0}}j\omega; \tag{A24}$$

## Appendix 3: Efficiency of the regenerator in harmonic regime

The enthalpy flux in the regenerator is :

$$H = c_g\rho_g S\overline{U}\overline{T}_g \tag{A25}$$

After developing

$$\dot{H} = c_g S\left(\xi\left(\rho_{g0} T_{g0} U_a\right) + \xi^2\left(\rho_{g0} T_{g0} U_{2a} + \rho_{g0} T_{ga} U_a + \rho_{ga} T_{g0} U_a\right) + ..\right)$$

considering temporal average values :

$$\langle U_a \rangle = \langle U_{a2} \rangle = \langle \rho_a U_a \rangle = 0 \qquad \text{(A26)}$$

$$\langle \dot{H} \rangle = \approx c_g \rho_{g0} S \frac{1}{2} \Re\left[T_{ga} \tilde{U}_a\right] = S \frac{1}{2} \Re\left[\tau\left(j\omega p_a \tilde{U}_a - c_g \rho_{g0} \frac{|U_a|^2}{\varepsilon} \frac{\partial T_{t0}}{\partial x}\right)\right] \qquad \text{(A27)}$$

So the regenerator efficiency is defined by eq. A17:

$$E = 1 - \frac{\langle \dot{H} \rangle|_L}{\langle \dot{W}_{acou} \rangle|_L} = 1 - \frac{\Re\left[\tau\left(j\omega p_{aL} \tilde{U}_{aL} - c_g \rho_{g0} \frac{|U_{aL}|^2}{\varepsilon} \frac{\partial T_{t0}}{\partial x}\Big|_L\right)\right]}{\Re\left[p_{aL} \tilde{U}_{aL}\right]} \qquad \text{(A28)}$$

where $\tilde{U}$ represent the complex conjugate.

## References.

| | | |
|---|---|---|
| compressor | volume ratio :<br>Dead volume ratio : | 1.31<br>0.197 |
| Gas tube | Length :<br>Volume ratio : | 70 mm<br>1 |
| regenerator | Length :<br>volume ratio :<br>Wire mesh porosity : | 55 mm<br>0.6963<br>0.707 |
| Heat exchangers | First exchanger volume ratio : 0.0655<br>Cold exchanger, volume ratio: 0.082<br>Warm exchanger, volume ratio : 0.0434 | |
| Capillaries | Compressor junction,<br><br>By-pass junction,<br><br>Reservoir junction, | Internal diameter : 1.57 mm<br>Length : 500 mm inox<br>Internal diameter : 0.7 mm<br>Length : 110 mm inox<br>Internal diameter : 0.7 mm<br>Length : 460 mm inox |

Tab. 1 Geometrical characteristics for the tested THALES CRYOGENIE DIPTR cooler .

Fig. 1 Equivalent electrical diagram for miniature DIPTR.

Fig. 2 Photography of the commercial DIPTR prototype (THALES CRYOGENIE) used for tests.

Fig. 3 Dimensionless pressure amplitude evolutions in pulse tube depending on $T_c/T_h$ ratio : experience, thermodynamic model and analogy model.

Fig. 4 Cooling power factor evolutions depending on $T_c/T_h$ ratio experience, thermodynamic model and analogy model.

Fig. 5 Reservoir capillary dimensionless parameter $Kk_r$ influence on refrigerating production ; temperatures ratio $T_c/T_h$ = 0.75 ; parameter : by-pass dimensionless parameter $Kk_1$.

Fig. 6 Velocities and pressure histories at the cold and warm extremities off the pulse tube

Fig.7 Cooling power evolutions depending on Tc :analogy model with and without the hypothesis of constant resistance and regenerator efficiency.

Fig.8 Histories of cold zone inlet temperature and pressure.

Fig.9 COP depending on compressor swept-volume

Fig.10 Net cooling production according to the regenerator length with constant volume.

Fig.11 Regenerator efficiency evolution versus regenerator length with constant volume.

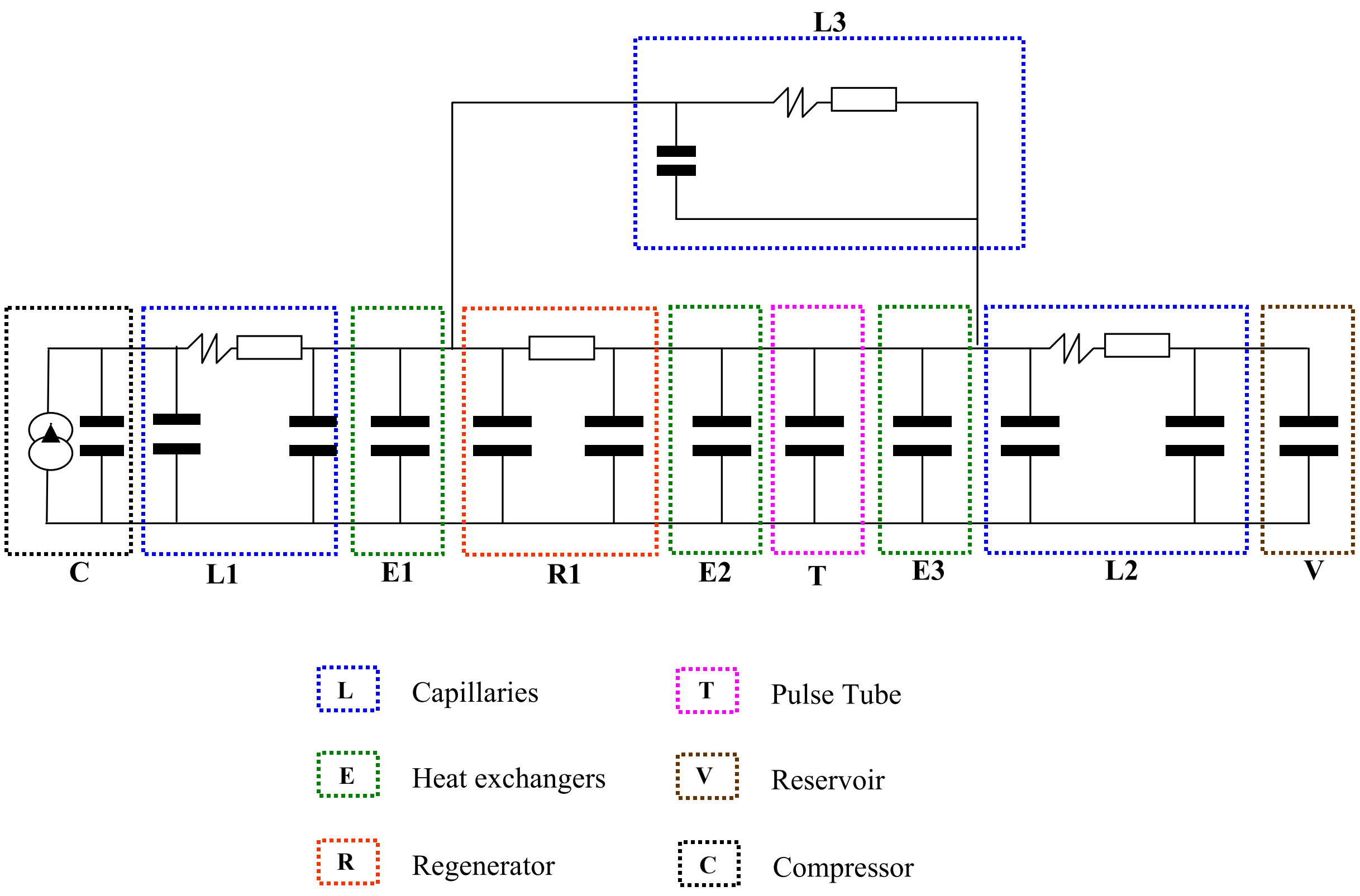

Fig. 1  Equivalent electrical diagram for miniature DIPTR.

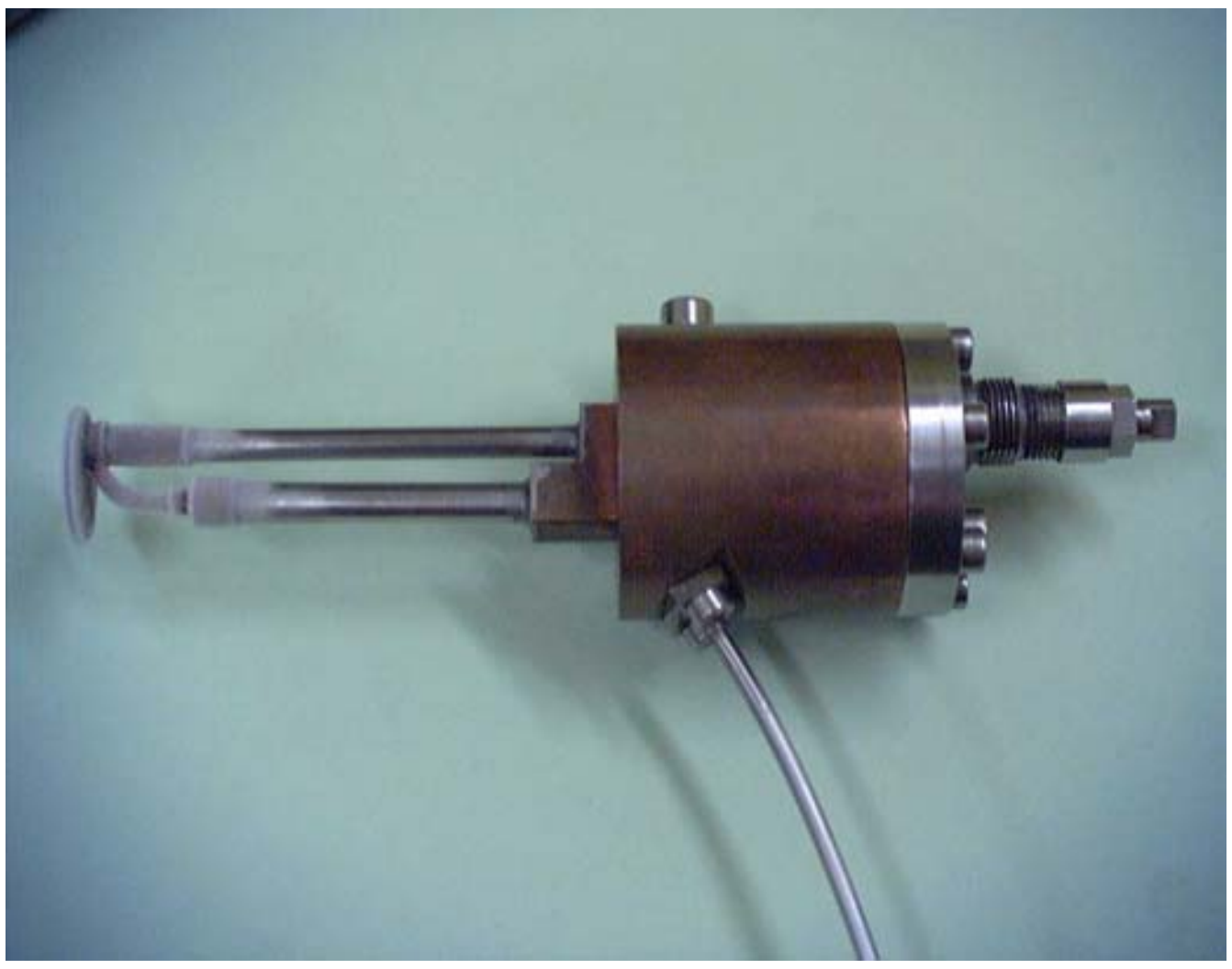

Fig. 2  Photography of the commercial DIPTR prototype  (THALES CRYOGENIE) used for tests.

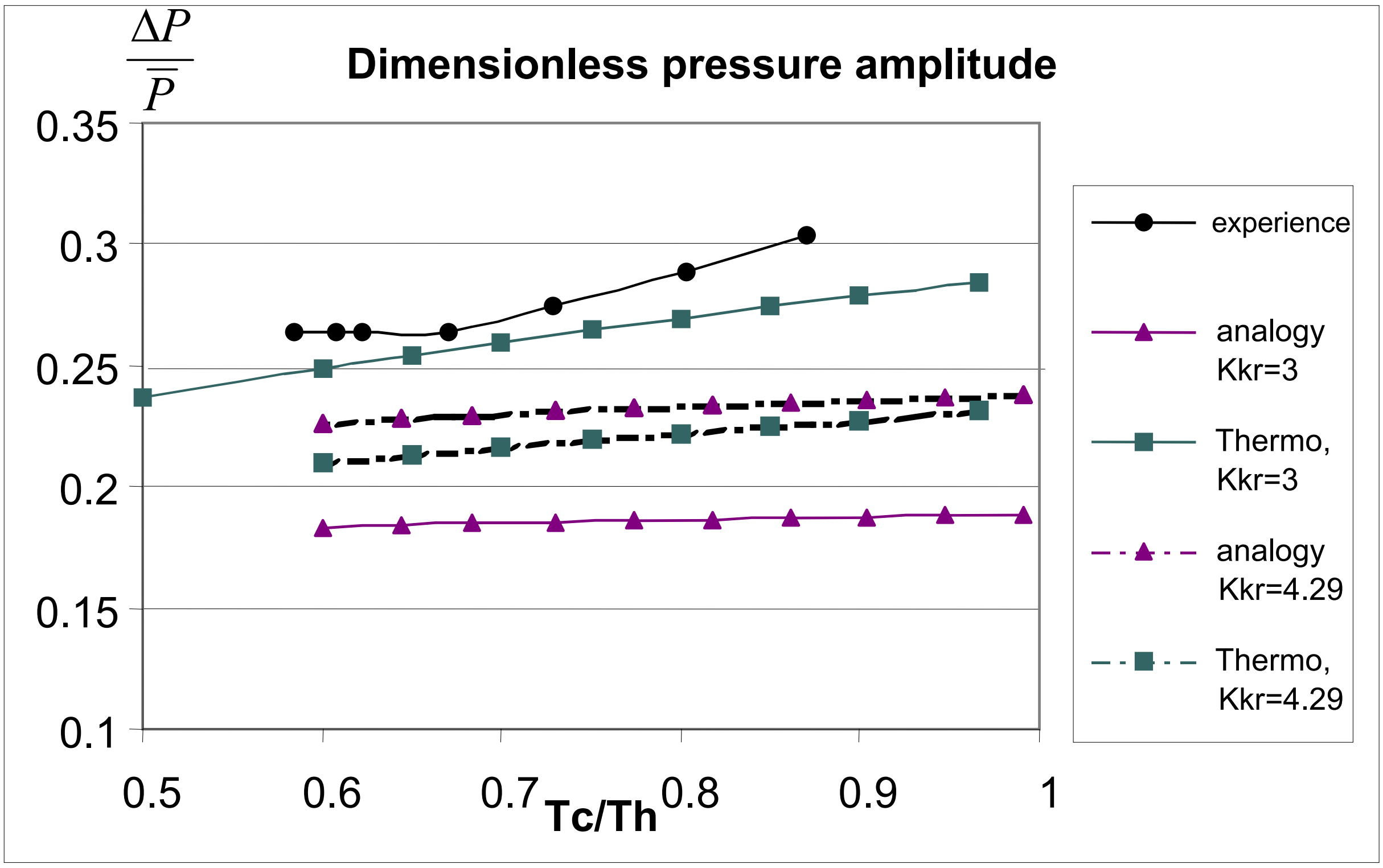

Fig. 3 Dimensionless pressure amplitude evolutions in pulse tube depending on $T_c/T_h$ ratio : experience, thermodynamic model and analogy model.

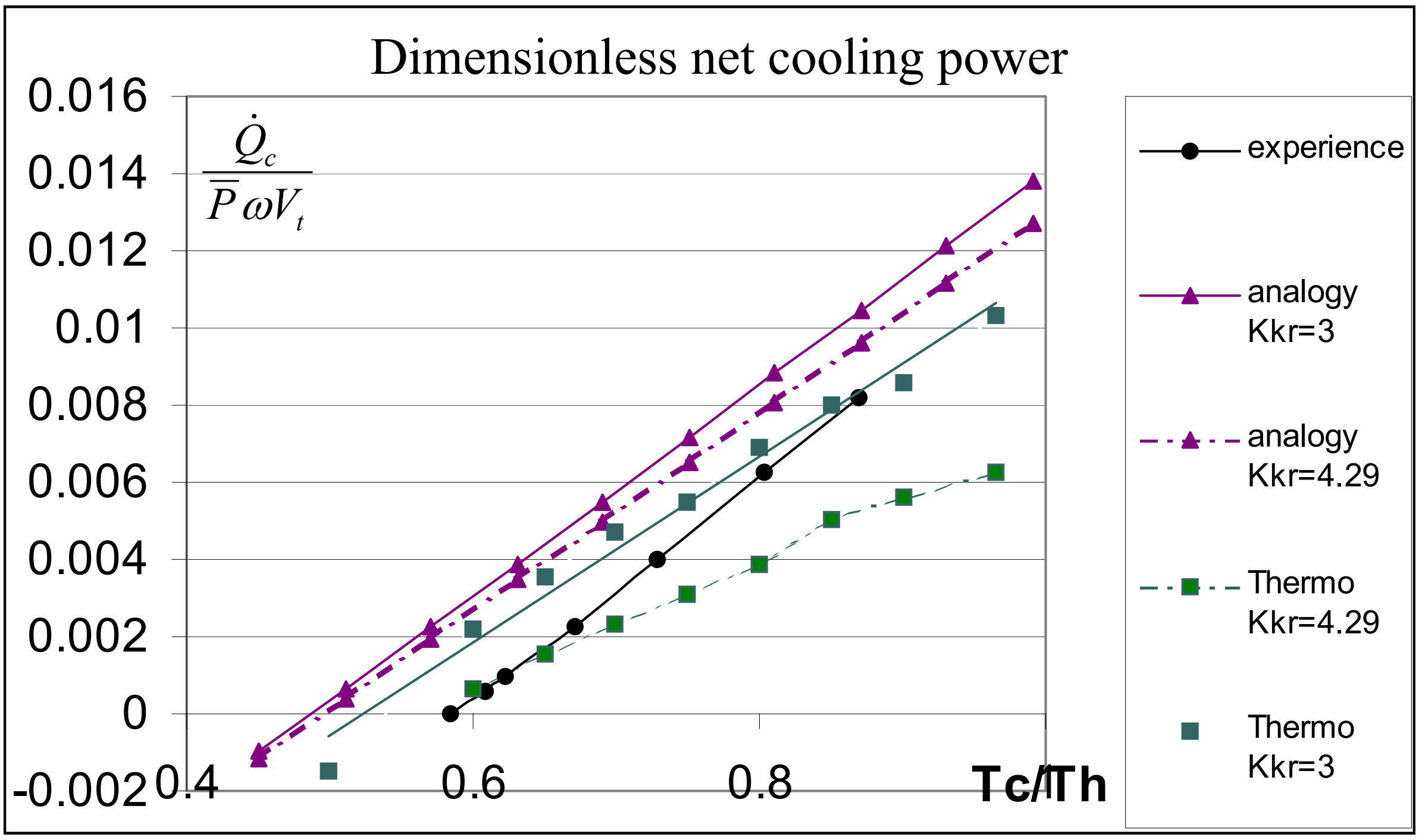

Fig. 4 Cooling power factor evolutions depending on $T_c/T_h$ ratio experience, thermodynamic model and analogy model.

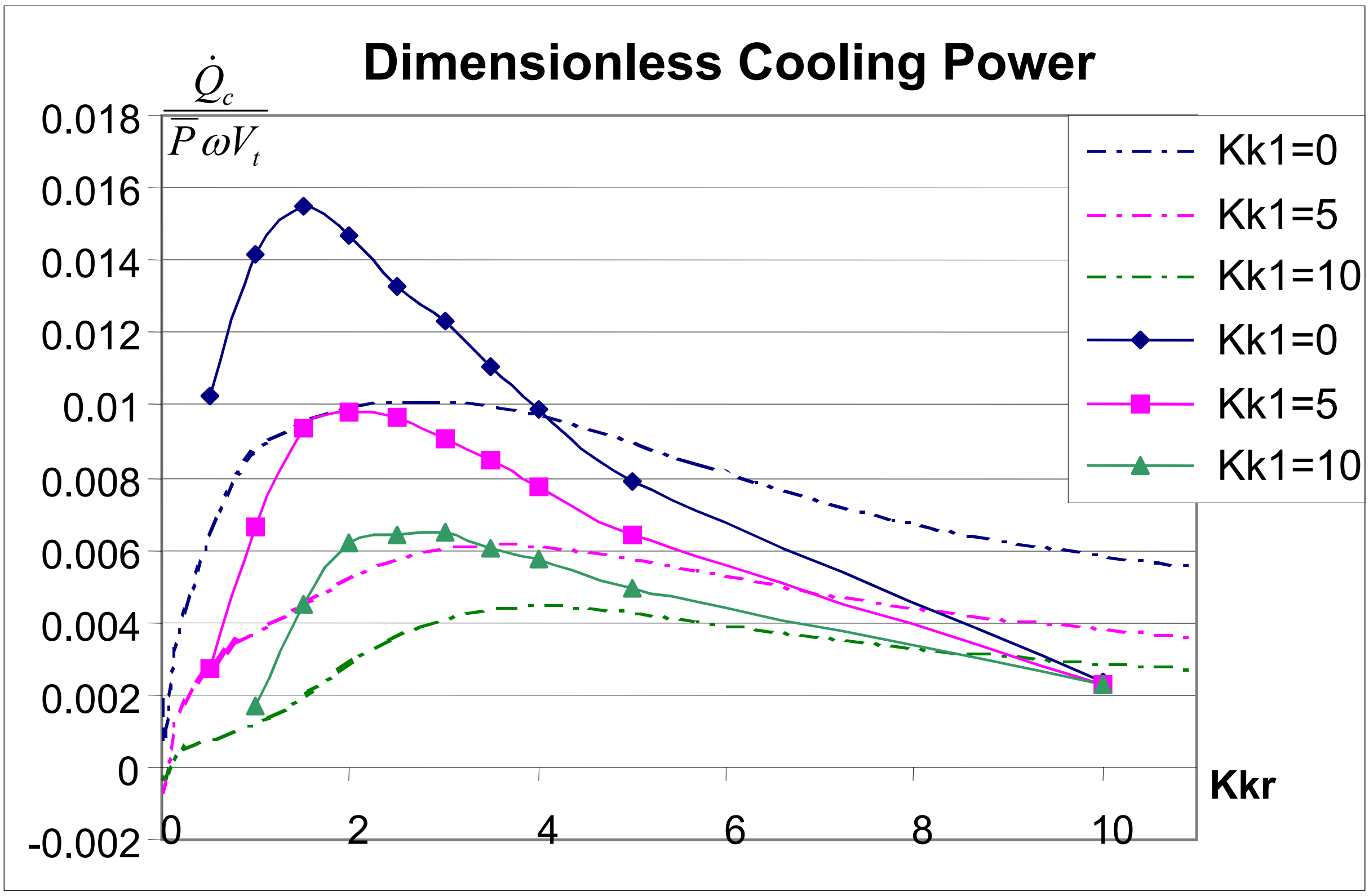

Fig. 5 Reservoir capillary dimensionless parameter $Kk_r$ influence on refrigerating production ; temperatures ratio $T_c / T_h$ = 0.75 ; parameter : by-pass dimensionless parameter $Kk_1$.

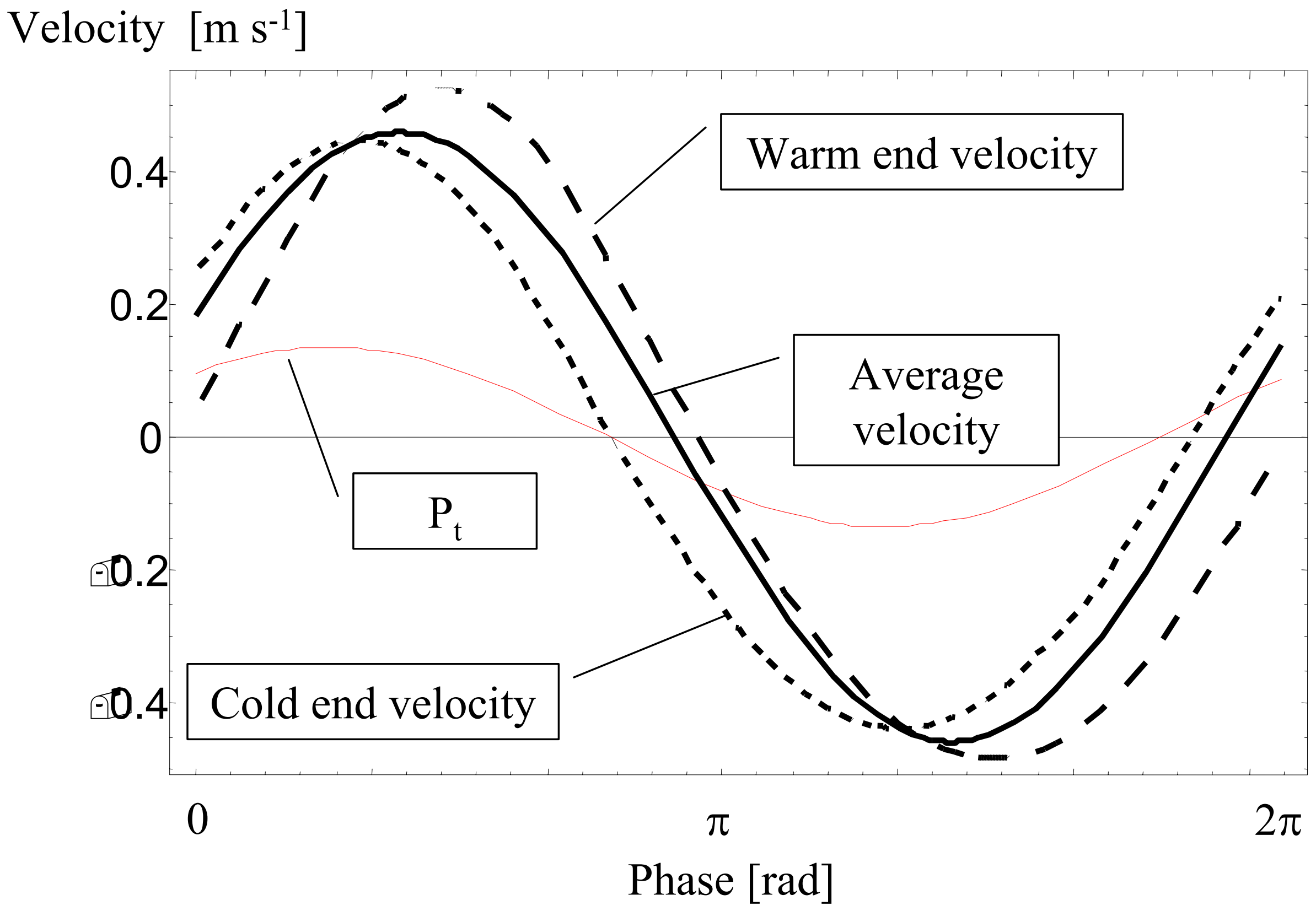

Fig. 6 Velocities and pressure histories at the cold and warm extremities off the pulse tube

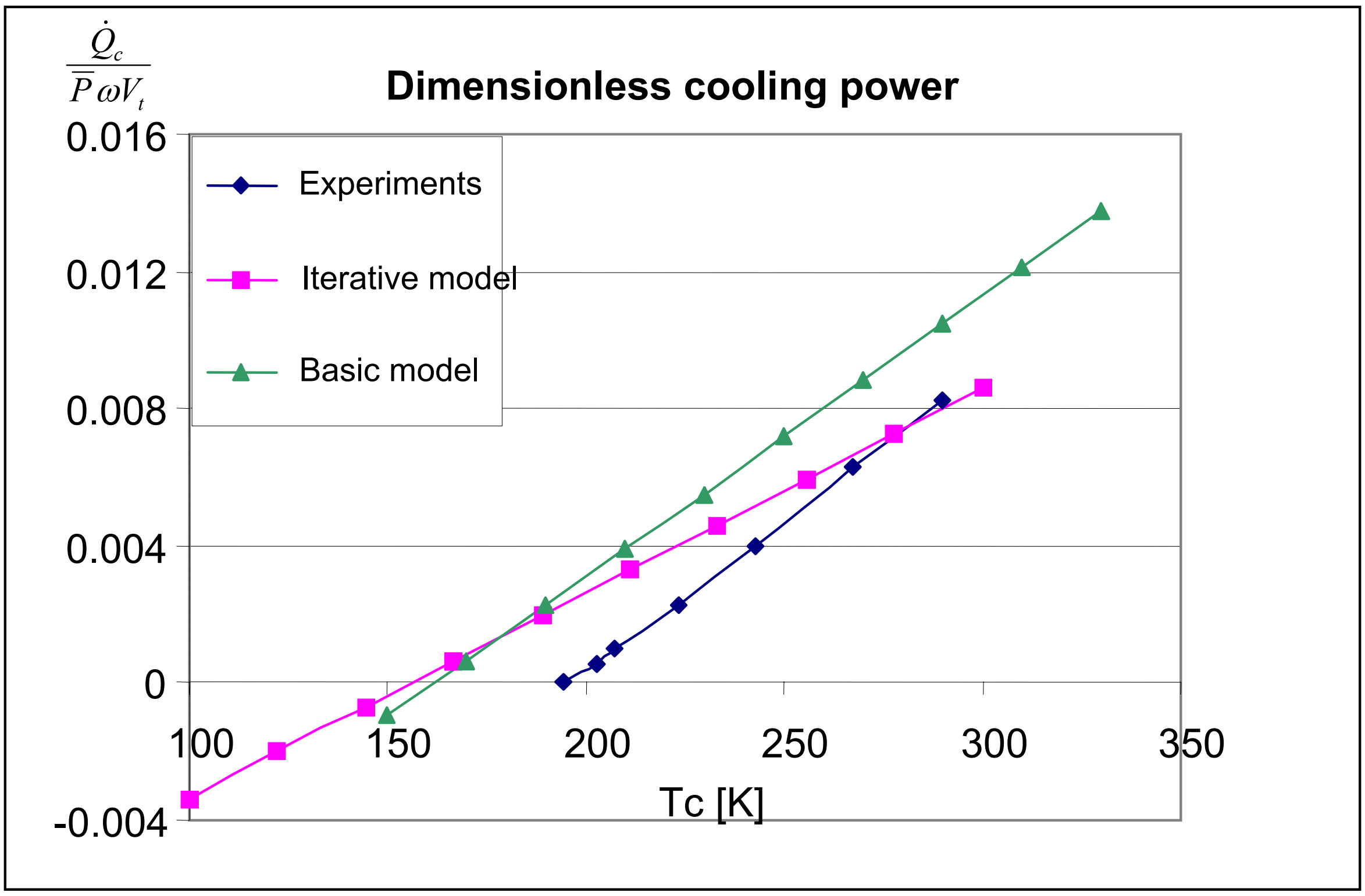

Fig.7 Cooling power evolutions depending on Tc :analogy model with and without the hypothesis of constant resistance and regenerator efficiency.

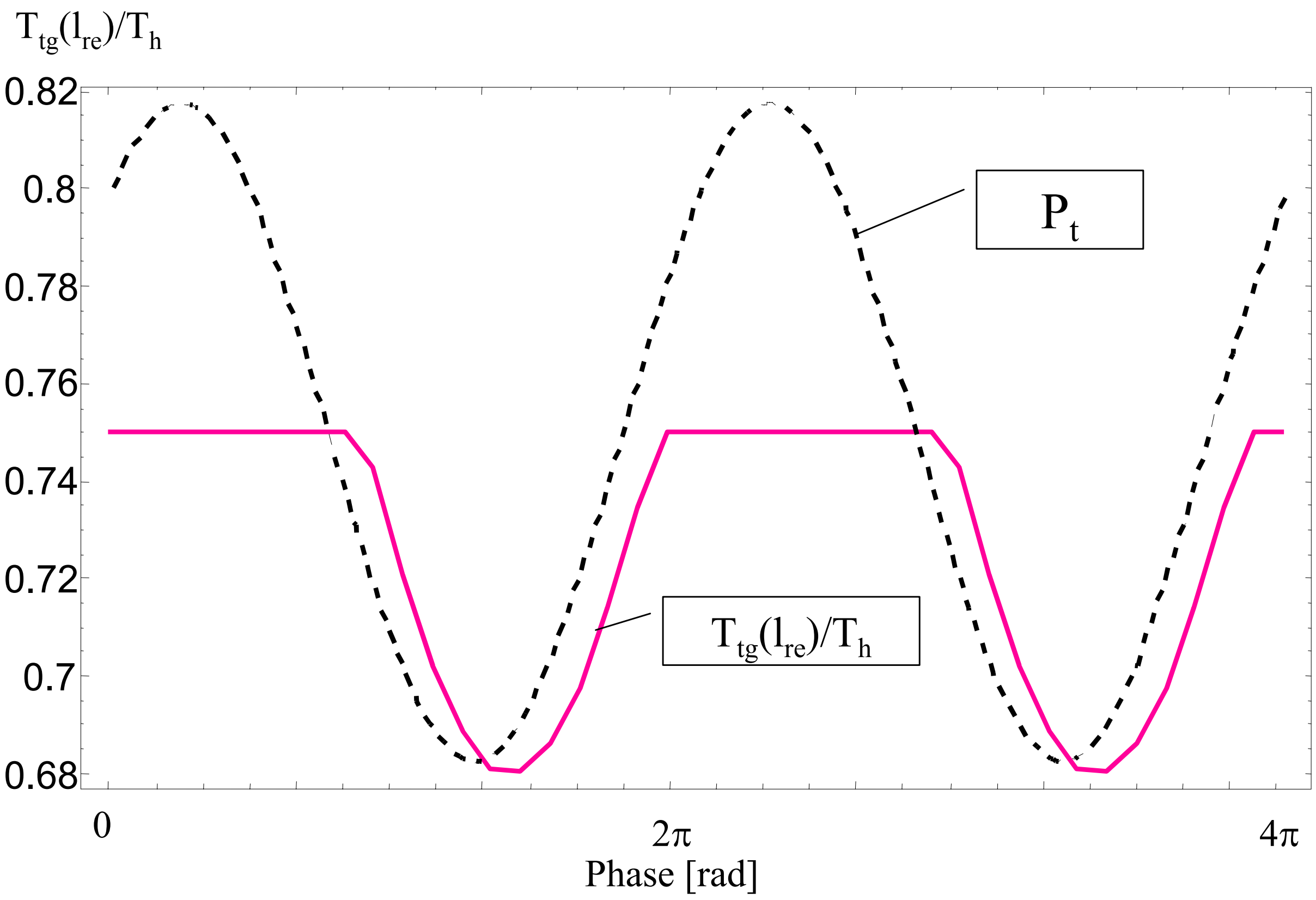

Fig.8 Histories of cold zone inlet temperature and pressure.

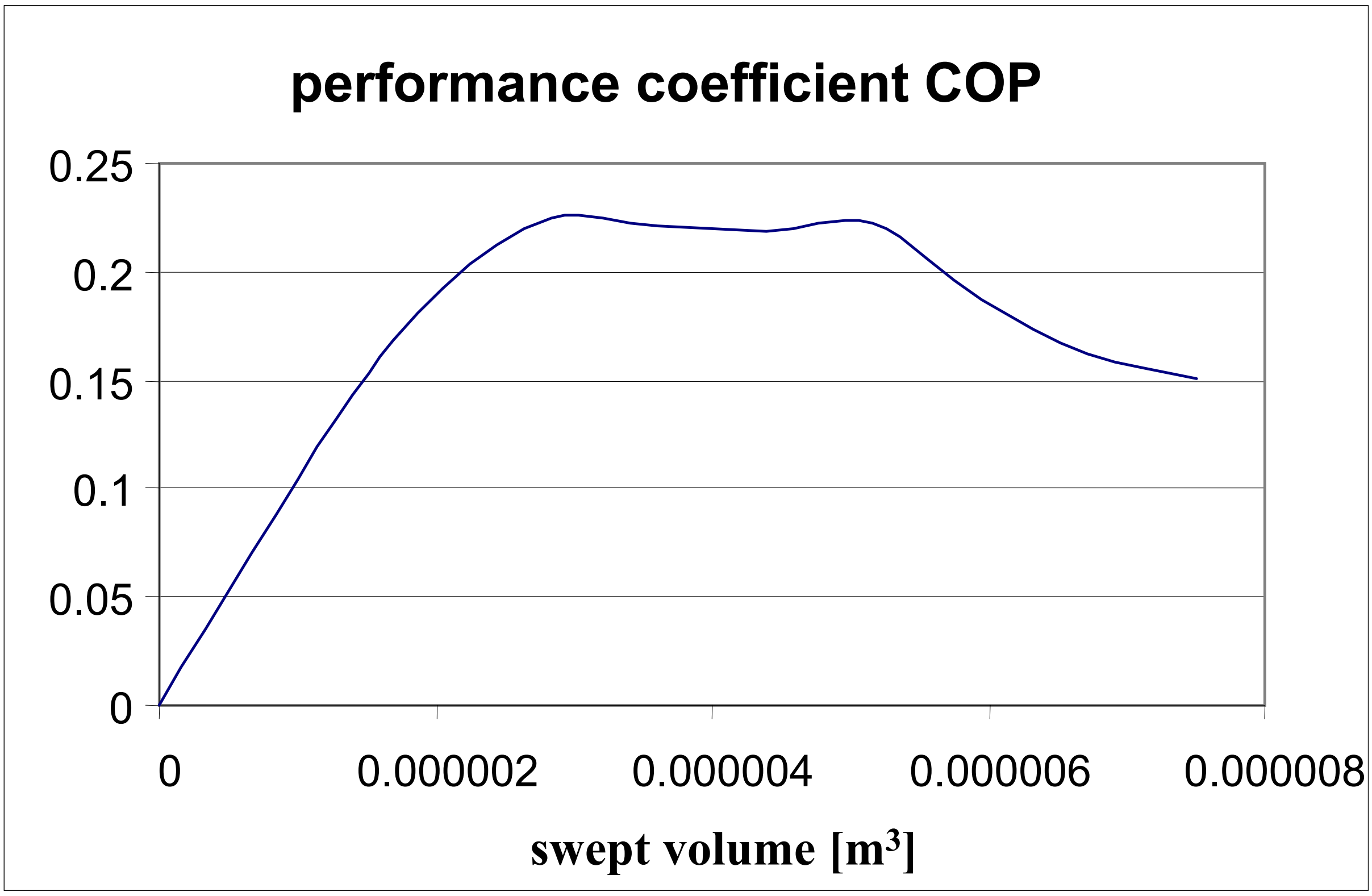

Fig.9 COP depending on compressor swept-volume

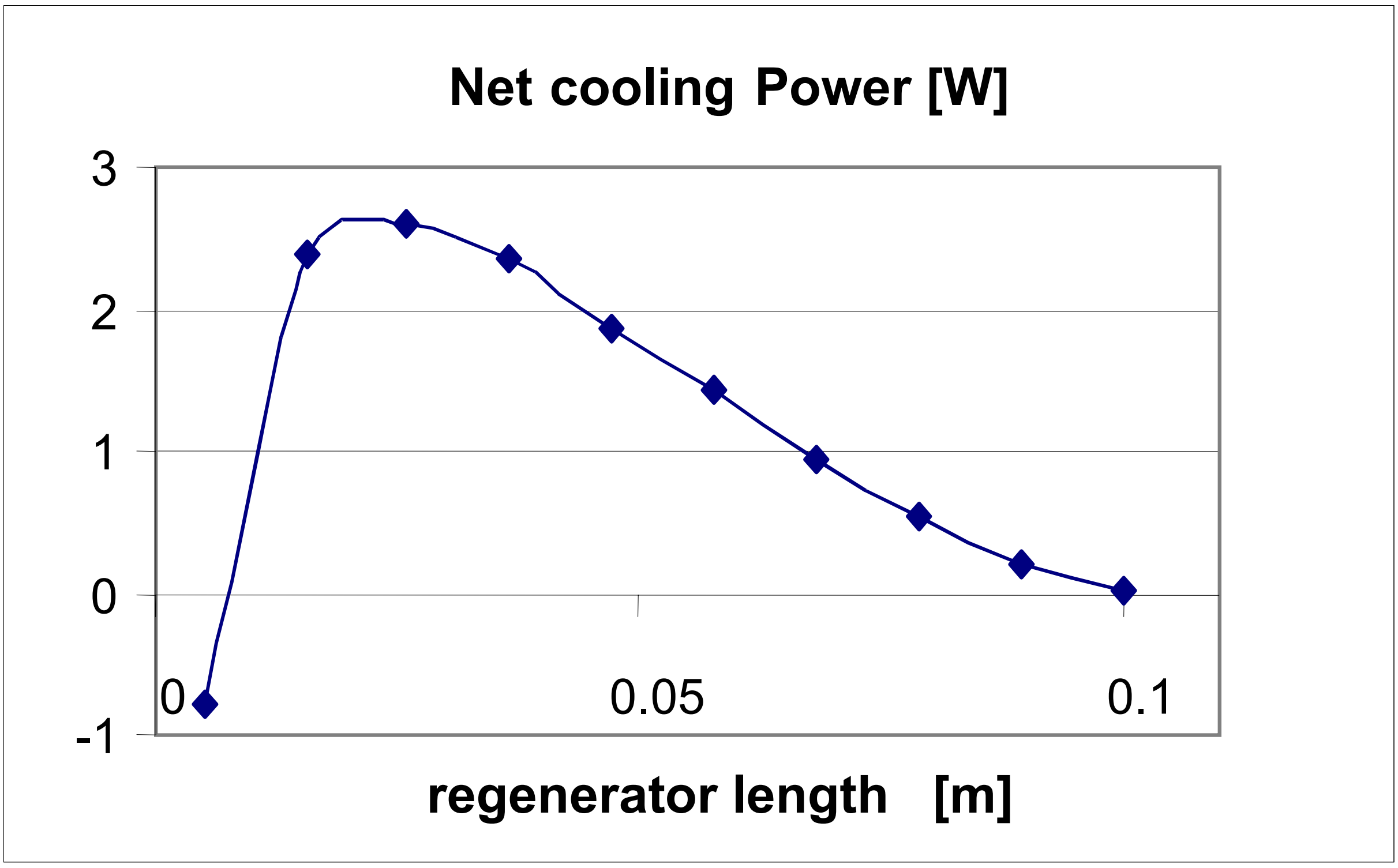

Fig.10 Net cooling production according to the regenerator length with constant volume.

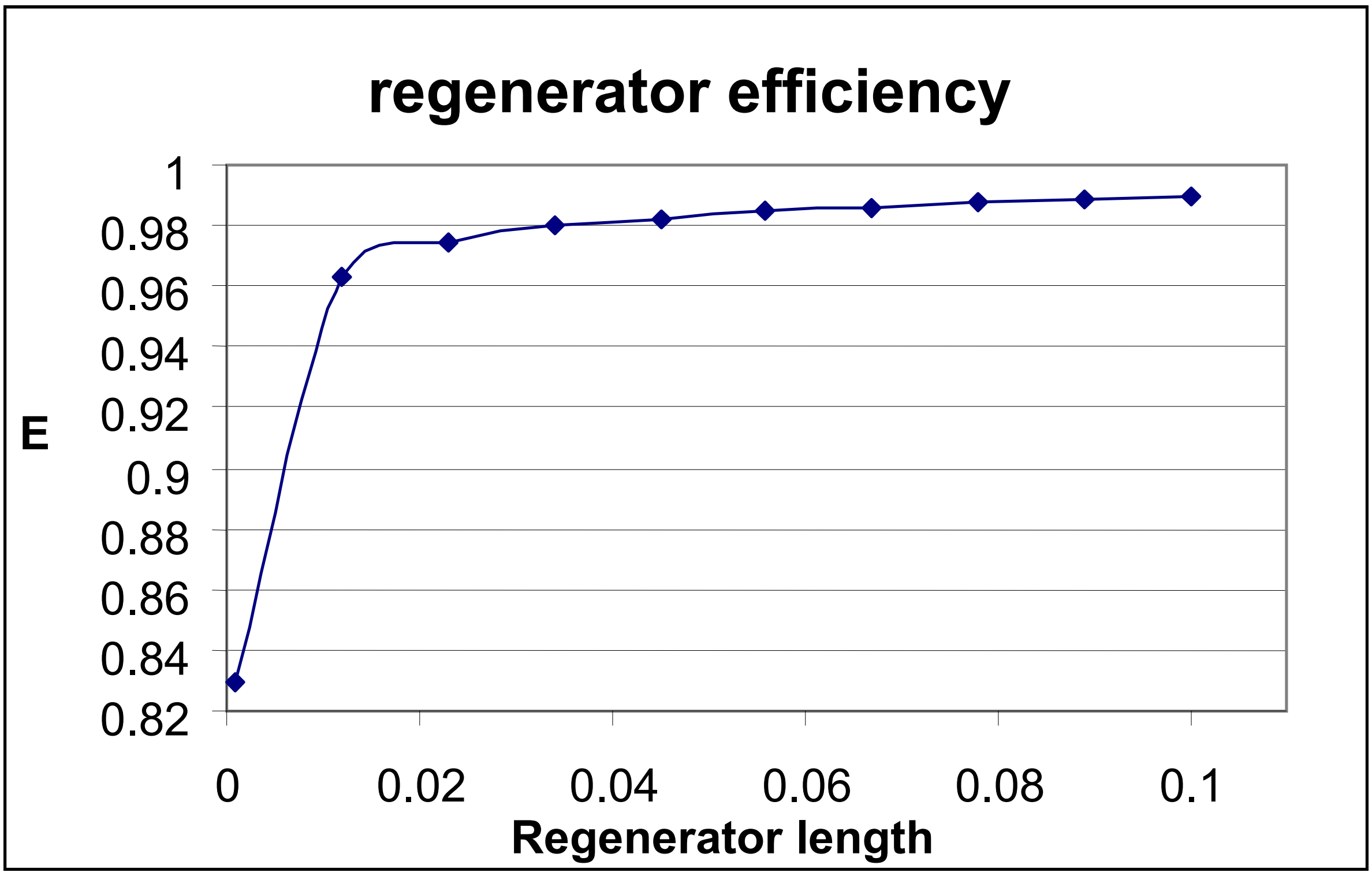

Fig.11 Regenerator efficiency evolution versus regenerator length with constant volume.